\documentclass[sigconf]{acmart}
\AtBeginDocument{%
  }

\copyrightyear{2025}
\acmYear{2025}
\setcopyright{acmlicensed}\acmConference[RecSys '25]{Proceedings of the Nineteenth ACM Conference on Recommender Systems}{September 22--26, 2025}{Prague, Czech Republic}
\acmBooktitle{Proceedings of the Nineteenth ACM Conference on Recommender Systems (RecSys '25), September 22--26, 2025, Prague, Czech Republic}
\acmDOI{10.1145/3705328.3748035}
\acmISBN{979-8-4007-1364-4/2025/09}

\usepackage{multirow}
\usepackage{enumitem}

\begin{document}

\title{On Inherited Popularity Bias in Cold-Start Item Recommendation}

\author{Gregor Meehan}
\email{gregor.meehan@qmul.ac.uk}
\orcid{0009-0007-2619-9299}
\affiliation{%
  \institution{Queen Mary University of London}
  \city{London}
  \country{United Kingdom}
}

\author{Johan Pauwels}
\email{j.pauwels@qmul.ac.uk}
\orcid{0000-0002-5805-7144}
\affiliation{%
  \institution{Queen Mary University of London}
  \city{London}
  \country{United Kingdom}
}

\renewcommand{\shortauthors}{Meehan and Pauwels}

\begin{abstract}
Collaborative filtering (CF) recommender systems struggle with making predictions on unseen, or `cold', items. Systems designed to address this challenge are often trained with supervision from warm CF models in order to leverage collaborative and content information from the available interaction data. However, since they learn to replicate the behavior of CF methods, cold-start models may therefore also learn to imitate their predictive biases. In this paper, we show that cold-start systems can inherit popularity bias, a common cause of recommender system unfairness arising when CF models overfit to more popular items, thereby maximizing user-oriented accuracy but neglecting rarer items. We demonstrate that cold-start recommenders not only mirror the popularity biases of warm models, but are in fact affected more severely: because they cannot infer popularity from interaction data, they instead attempt to estimate it based solely on content features. This leads to significant over-prediction of certain cold items with similar content to popular warm items, even if their ground truth popularity is very low. Through experiments on three multimedia datasets, we analyze the impact of this behavior on three generative cold-start methods. We then describe a simple post-processing bias mitigation method that, by using embedding magnitude as a proxy for predicted popularity, can produce more balanced recommendations with limited harm to user-oriented cold-start accuracy.  
\end{abstract}


\maketitle

\section{Introduction} \label{sec:intro}
Recommender systems are crucial to the user experience in the modern digital ecosystem. To provide personalized suggestions, these models are trained using historical user-item interactions, typically learning unique representations of each user or item's identity (ID) and predicting user-item preferences via similarity in the resulting embeddings. However, such models face the cold-start problem ~\cite{schein2002methods}: since newly added users or items do not have associated ID embeddings, predictions cannot be made about them.

In this paper, we focus on the item variant of this problem, which is typically addressed by using item content features to augment model predictions. Some item cold-start methods~\cite{volkovs2017dropoutnet, CLCRec, zhou2023contrastive, NEMCF, kim2024content, wang2024preference} train models from scratch to make predictions with user ID embeddings and encoded item content alone. However, this burden on the user embeddings can significantly degrade prediction quality for warm items~\cite{huang_aligning_2023}. An alternative approach is generative, using pre-trained collaborative filtering (CF) models to provide supervision to content encoders that generate synthetic ID embeddings. The encoders are trained to either minimize distances between encoded content and pre-trained item embeddings~\cite{van2013deep, barkan2019cb2cf, zhu_recommendation_2020, du2020learn, chen2022generative, bai2023gorec} or to reproduce CF model ranking behavior~\cite{chen2022generative, du2020learn, huang_aligning_2023}. The generated embeddings for cold items can then be used for prediction, while warm item predictions can leverage the pre-trained CF embeddings, meaning there is no compromise in warm item performance. 

However, due to imbalances in training data and parameter tuning regimes focused on maximizing accuracy, CF-based recommender systems are prone to severe predictive biases relating to item popularity \cite{ahanger2022popularity, klimashevskaia2024survey}. These biased models suggest the most popular items to users even more than their popularity would warrant, while less popular items are neglected \cite{mansoury2020feedback}. This behavior can lead to unfair outcomes not just for rarer items~\cite{abdollahpouri2020connection, deldjoo2024fairness} but for all recommender system stakeholders~\cite{abdollahpouri2020multistakeholder}. 

Although there has been much work in mitigating this popularity bias for warm items \cite{klimashevskaia2024survey}, few studies~\cite{zhu2021fairness} consider its implications in a cold-start context. In this work, we examine how CF predictive biases affect generative cold-start recommenders, demonstrating that generative models learn to imitate their supervisory warm model and overexpose certain items, i.e.\ to map the content features of these items to high preference scores for a large portion of the user population. The cold-start models then replicate this pattern for new items during inference, leading to some cold items being treated as very popular without any direct evidence from actual user interactions. These inherited behaviors therefore not only decrease item fairness but can also reduce overall recommendation quality. To address this issue, we propose a post-processing mitigation approach that balances the distribution of predictions across items by rescaling the magnitudes of learned item representations. Through experiments with three generative models across three real-world multimedia datasets, we show that this simple mitigation step can improve item fairness with minimal harm to user-oriented accuracy. We provide code and other resources to reproduce our results at \href{https://github.com/gmeehan96/Cold-PopBias}{https://github.com/gmeehan96/Cold-PopBias}.

\section{Related Work}
Research on item popularity bias in warm item recommendation typically aims to both measure the degree of bias and to mitigate its impact on item fairness~\cite{klimashevskaia2024survey}. Popularity bias can be measured by the coverage~\cite{oh2011novel} or rate~\cite{abdollahpouri2017controlling, abdollahpouri2019managing} of long-tail items in a system's predictions, quantifying to what extent rarer items are exposed to users. This analysis is sometimes narrowed to only items of interest to a user~\cite{boratto2021connecting, zhu2021popularity}, i.e.\ measuring whether a system consistently ranks long-tail items lower than popular ones even when a user has interacted with both.

Many mitigation methods reduce bias by adjusting model training, adding loss terms to penalize overreliance on popularity for prediction~\cite{abdollahpouri2017controlling,wei2021model,zhu2021popularity, boratto2021connecting, rhee2022countering} or to increase alignment between short-head and long-tail items~\cite{cai2024mitigating,cai2024popularity,li2024popularity}. Other works use pseudo-labelling on knowledge graphs \cite{togashi2021alleviating} or take a causal approach~\cite{wei2021model, zhang2024robust} to isolate the effect of popularity bias so that it can be removed. By contrast, post-processing methods change model predictions after training is complete to promote long-tail items, either by simple adjustment of item rankings~\cite{abdollahpouri2021user, zhu2021popularity} or with more complex algorithmic approaches~\cite{eskandanian2020using, mansoury2020fairmatch, zhu2021fairness}.

Although some works~\cite{wei2022comprehensive, zhang2024enhancing} look at fairness in meta-learning approaches for cold user recommendation, similar research in the cold item context is limited. One notable exception ~\cite{zhu2021fairness} describes a framework for evaluating fairness among new items, and proposes a learnable post-processing bias mitigation method that uses autoencoders to align preference score distributions across both users and items.

Although related, our study differs from~\cite{zhu2021fairness} in several ways. Firstly, we focus specifically on popularity-related fairness, and examine more closely how cold-start models develop and manifest popularity biases. We also augment our evaluation framework by measuring fairness in item exposure as well as in accuracy. Finally, our post-processing mitigation method is more flexible than the autoencoder proposed in \cite{zhu2021fairness}, which requires the calculation of user-item score vectors across the entire user population before it can be trained, a potentially impracticable requirement in large-scale industrial settings. Furthermore, since these score vectors are fixed in size, the model will not be able to make predictions for new users without retraining, negating a key practical benefit of post-processing approaches. By contrast, our method requires no training and only relies on statistics of the generated item embeddings, meaning that it can be directly applied for new users.

\section{Experimental Framework}
We begin by introducing the datasets and cold-start models considered in our analysis, as well as our evaluation framework.

\subsection{Data}
We consider three multimedia datasets from the MMRec toolbox~\cite{zhou2023mmrec}: micro-video dataset \textbf{Microlens}~\cite{ni2023content} and the \textbf{Clothing} and \textbf{Electronics} Amazon E-commerce subsets~\cite{mcauley2015image}.  Dataset statistics and data mode information are in Table \ref{table:data_stats}. To create a single input feature vector for each dataset, we L2-normalize the content vectors from each data mode and concatenate them. We exclude the image features for Clothing because, as noted in other works~\cite{zhang_modality-balanced_2024, malitesta2023popularity}, we find that including them lowers performance. Following previous cold-start studies~\cite{chen2022generative, huang_aligning_2023} we split the datasets twice to facilitate training of the cold models and of the warm CF model. First, we divide the items 80:20 into warm and cold sets, with the cold set split 50:50 into validation and test. We then split the warm item user interactions 80:10:10 into training, validation, and test sets.
\begin{table}[]
    \caption{Statistics for datasets used. The size of the feature vector for each data mode is indicated in parentheses. \label{table:data_stats}}
  \small
  \begin{tabular}{cccccc}
  \toprule
  Dataset                 & Users    & Items & Interactions & Density & Modes (dim)   \\ \midrule
  Clothing   & 39,387    & 23,033 & 278,677 & 0.031\% & Text (384) \\[1pt] \\[-0.85em]
  \multirow{2}{*}{Electronics} & \multirow{2}{*}{192,403}  & \multirow{2}{*}{63,001} & \multirow{2}{*}{1,689,188} & \multirow{2}{*}{0.014\%} & Image (4096) \\ 
  &                          &                         &                            &                          & Text (384) \\[1pt] \\[-0.85em]  
  \multirow{3}{*}{Microlens} & \multirow{3}{*}{98,129}  & \multirow{3}{*}{17,228} & \multirow{3}{*}{705,174} & \multirow{3}{*}{0.042\%} & Image (1024) \\
  &                          &                         &                            &                          & Text (1024) \\
   &                          &                         &                            &                          & Video (768) 
  \\ \bottomrule
  \end{tabular}

  \end{table}

\subsection{Models}
We analyze three generative cold-start models: \textbf{Heater} \cite{zhu_recommendation_2020} uses mixtures-of-experts \cite{shazeer2017outrageously} to transform item content features, and randomly swaps between item ID embeddings and projected content during training.  \textbf{GAR}~\cite{chen2022generative} uses generative-adversarial training between the pre-trained CF model and a content-based generator, while \textbf{GoRec}~\cite{bai2023gorec} trains a conditional variational autoencoder (CVAE) \cite{sohn2015learning} to reconstruct item ID embeddings conditioned on content features. We implement these models using their publicly available repositories, and replicate typical cold-start hyperparameter tuning by aiming to maximize cold validation set accuracy without concern for bias.

For the pre-trained CF model, we choose multimodal recommender \textbf{FREEDOM}~\cite{zhou_tale_2023}, which, unlike pure CF methods, is able to leverage item content features. FREEDOM's hyperparameters are also tuned based on validation accuracy. Finally, we include non-parametric baseline \textbf{KNN} \cite{sedhain2014social}, which calculates preference scores by content similarity to a user's interacted items.

\begin{figure*}[]
\begin{tabular}{cc}
     \includegraphics[width=100mm,trim={0 0.75cm 0 0}]{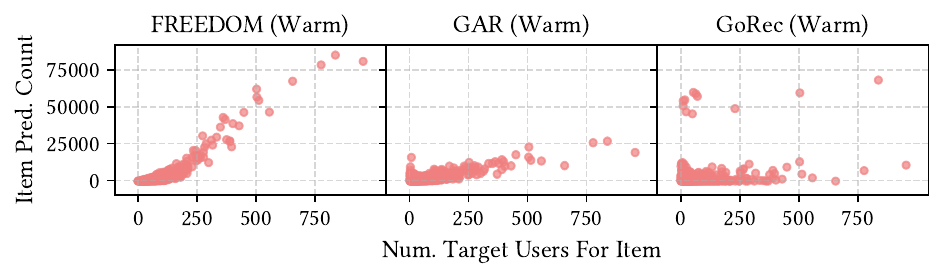} & \includegraphics[width=73mm,trim={0 0.75cm 0 0}]{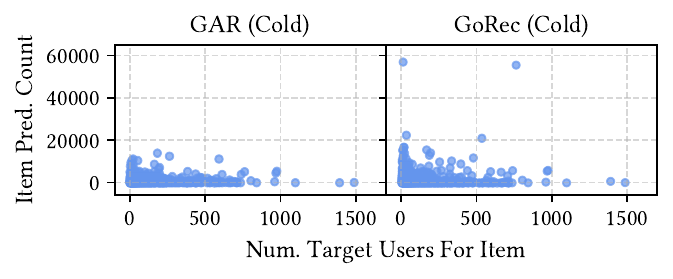} 

\end{tabular}
\caption{Warm and cold item prediction counts with $k=20$ against the number of target users (i.e. the number of times an item appears in the validation or test set interactions) for the Electronics dataset. Each dot represents an item: we only include items with at least one holdout set interaction, so there are 44,083 in the warm plots and 12,601 in the cold plots.} \label{full_dists}
\Description{Scatter plots showing prediction counts by number of users per item. This figure consists of five scatter plots, each showing individual items plotted by the number of users who interacted with the item (x-axis, labeled “Num. Target Users For Item”) and the number of times the item was predicted by a model (y-axis, labeled “Item Pred. Count”). The three plots on the left correspond to warm items from FREEDOM, GAR, and GoRec. The two plots on the right correspond to cold items from GAR and GoRec models. In the warm item plots for FREEDOM and GAR, most items are positioned toward the lower left, where both target user count and prediction count are low, with a gradual increase in values toward the upper right. In FREEDOM, the largest prediction counts range up to about 85000, while for GAR the maximum prediction count is just above 25000. For GoRec, there is a less clear increase in prediction counts, and not much correlation: while some items have prediction counts above 50000, many of these occur for items with low target users (less than 100). In the GAR cold item plot, the maximum prediction count is less than 20000, and there is no trend in the number of target users for when the prediction count increases. This is also true for the cold GoRec plot, with a larger spread in prediction counts, as the maximum count is approximately 60000. }
\end{figure*}

\begin{figure}[]
\begin{tabular}{cc}
     \includegraphics[width=40mm,trim={0 0.75cm 0 0}]{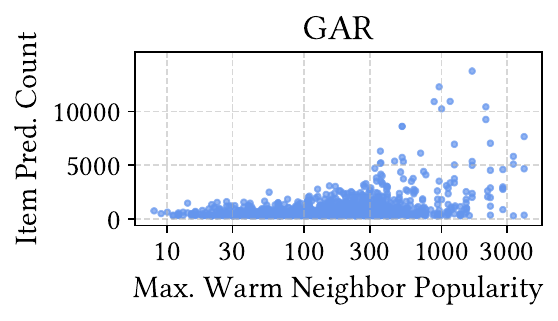} & \includegraphics[width=40mm,trim={0 0.75cm 0 0}]{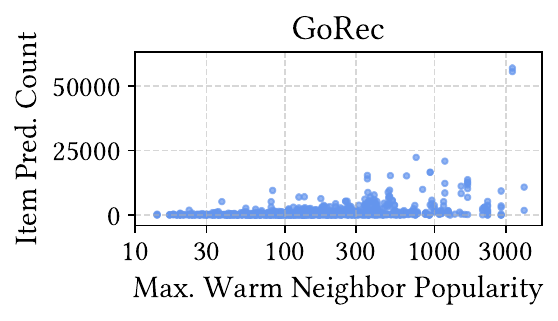} 

\end{tabular}
\caption{Cold item prediction counts for the top 10\% most predicted cold Electronics items. The x-axis is the maximum popularity value among the top 10 closest warm neighbors to each item by cosine similarity in the content features.} \label{neighbor_pops}
\Description{Scatter plots of cold item prediction counts versus warm neighbor popularity. This figure contains two side-by-side scatter plots. The left plot is labeled “GAR” and the right plot is labeled “GoRec.” Both plots have the same axes: the x-axis is labeled “Max. Warm Neighbor Popularity,” and the y-axis is labeled “Item Pred. Count.” The x-axis ranges from approximately 0 to slightly above 3000 and uses a logarithmic scale. The y-axis ranges from 0 to just above 10,000 in the GAR plot and from 0 to just above 50,000 in the GoRec plot; it uses a linear scale. In the GAR plot, most points are concentrated in the bottom third, particularly below 1000 on the x-axis and below 2000 on the y-axis. A small number of outliers appear higher on the y-axis, with prediction counts exceeding 5000 and neighbor popularity extending past 300. In the GoRec plots, the structure is similar, except the outliers have much higher prediction counts: many range up to 25000 and two are over 50000.}
\end{figure}

\subsection{Evaluation}\label{sec:eval}
We use ranking metrics \textbf{NDCG@$k$ }and \textbf{Recall@$k$} to measure user-oriented cold-start performance. Following \cite{zhu2021fairness}, we quantify item fairness by Max-Min Opportunity \cite{rawls2001justice}, i.e.\ by the outcomes for the worst-off items. As in \cite{zhu2021fairness}, we measure these outcomes with Mean Discounted Gain (MDG@$k$), an item-oriented equivalent of NDCG. Since our datasets are highly sparse (e.g.\ Electronics has an interaction density of 0.014\%), many items are not suggested to any of their matched users, so when measuring low-end performance we average across the bottom 80\% of items\footnote{We choose 80\% because this value is often used to designate the `long-tail' in a popularity bias context, e.g. in \cite{cai2024popularity, lee2024post}.} by MDG (\textbf{MDG-Min80\%}) to keep the metric informative. As in \cite{zhu2021fairness} we also measure best-served (\textbf{MDG-Max5\%}) and overall (\textbf{MDG-All}) item performance. 

Beyond item-oriented accuracy, we also evaluate fairness in item exposure levels. Typically this quantifies the extent to which less popular items appear in the model's top suggestions \cite{abdollahpouri2017controlling, abdollahpouri2019managing}. However, cold item popularity is not known to the model, so we instead quantify exposure unfairness using \textbf{Gini-Diversity} \cite{adomavicius2011improving}. This metric measures distributional equality in item prediction counts, with higher values indicating that appearances in the top-$k$ are spread more evenly among the item population.

\section{Inherited Popularity Bias}
In this section, we first show that generative cold-start models inherit biased predictive patterns from their supervisory warm model, then propose a post-processing method to mitigate the resulting impact on item fairness. We leave to future work the analysis of how applying existing debiasing methods to the warm model would affect downstream cold-start fairness. 

\subsection{Motivation}\label{sec:motiv}

We motivate our work with an in-depth study of predictive behavior in generative cold-start models. We conduct this study on the Electronics dataset, but note that Clothing and Microlens exhibit similar trends, and include analyses of these datasets in the supplementary document in our repository. We also omit Heater, since its behavior is very similar to GAR's. We visualize in Figure \ref{full_dists} how frequently each item appears in the top 20 ranking positions for FREEDOM, GAR, and GoRec across the Electronics warm and cold holdout sets, plotted against the number of users who interact with that item in the holdout sets. Here the warm item predictions for GAR and GoRec are made using only the item content (i.e.\ treating them as cold items). We also note that the warm holdout set interactions are selected by independent sampling, so the number of target users for a warm item is correlated with its training set popularity.  

We observe first that FREEDOM's warm item behavior shows clear signs of popularity bias. Most of the 44,083 warm items have low target users and correspondingly low prediction counts, while a small subset of overexposed items are recommended far more than their popularity would merit. We can see in Figure \ref{full_dists} that GAR and GoRec also heavily favor a small set of items in both the warm and cold item sets.\footnote{GoRec's largest warm item prediction counts do not always occur for the most popular items due to its method for content-based inference, which uses cluster-based content similarity to infer approximate ID embedding inputs to its CVAE, and therefore cannot rely on a specific item's training set popularity.} Because they are trained using FREEDOM's predictions, these models learn to map the content features of a select few items to ID embeddings that will result in high rankings for large numbers of users. This behavior leads to severely imbalanced recommendations: for example, GoRec's top 50 items by prediction count make up 27.6\% of all top-20 user rankings across the 12,601 cold items, while 1,326 items do not appear in the top 20 for any user, clearly violating Max-Min fairness as discussed in Section \ref{sec:eval}.  

\begin{figure}[]
\begin{tabular}{cc}
      \includegraphics[width=40mm,trim={0 0.75cm 0 0}]{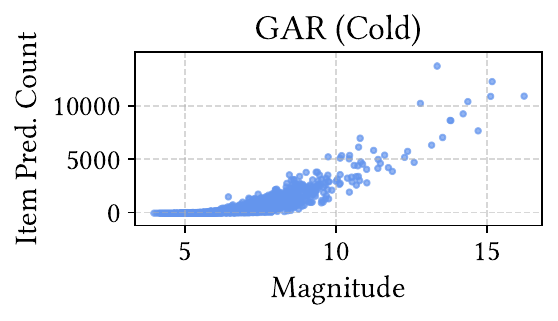} & \includegraphics[width=40mm,trim={0 0.75cm 0 0}]{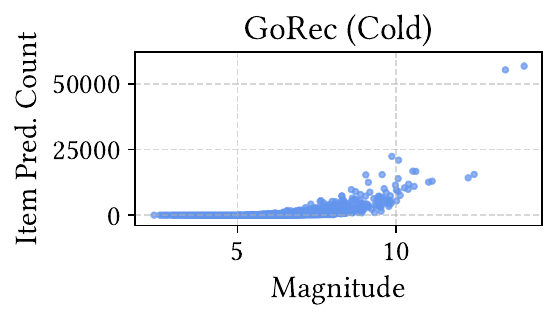} \\ 

\end{tabular}
\caption{Cold prediction counts  at $k=20$ against item vector magnitude in the Electronics dataset.} \label{fig:mag}
\Description{Scatter plots of cold item prediction counts vs. vector magnitude. This figure contains two scatter plots, one for the GAR model and one for the GoRec model, each visualizing cold items. The x-axis shows the magnitude of the item’s learned embedding vector. The y-axis shows the number of times each cold item was predicted. Most points are clustered within a magnitude range of 5 to 15 on the x-axis. In the GAR plot, prediction counts extend up to around 17,500, and points are more broadly distributed along both axes. Several points appear above 10,000 predictions and are located toward the right side of the plot, where magnitudes are highest. The GoRec plot shows a similar structure, with the extreme values in the top right having a higher prediction count (ranging to above 50000).}
\end{figure}

Furthermore, since these models have no ground truth popularity information for cold items, they are effectively estimating how popular new items will be based on their content features. Figure~\ref{neighbor_pops} shows that overexposed cold items are often close neighbors by content similarity to very popular warm items (typically, in the top 1\% by popularity); we provide statistical evidence for this claim in the supplementary material. Due to bias in their training, the models learn to associate certain content feature patterns with very high levels of exposure, so new items with similar features are often also assumed to be popular. However, the cold item graphs in Figure \ref{full_dists} show that this assumption is frequently wrong, and can result in tens of thousands of top ranking spots being dedicated to items with only a handful of interested users, while truly popular items are often neglected. 

\begin{table*}[!h]
\caption{User and item-oriented accuracy metrics along with exposure-based diversity across all datasets and models. Values where magnitude scaling (MS) provides a statistically significant (p < 0.01) improvement in performance over the base models are marked with asterisks ($^\ast$), while statistically significant losses in performance are marked with daggers ($^\dagger$). Values are bolded where MS improves performance by at least 10\%.}\label{tab:full_results} 
\small
\begin{tabular}{cc@{\hskip 7pt}l@{\hskip 6pt}l@{\hskip 15pt}l@{\hskip 5pt}l@{\hskip 6pt}l@{\hskip 10pt}lc@{\hskip 3pt} l@{\hskip 6pt}l@{\hskip 15pt}l@{\hskip 5pt}l@{\hskip 6pt}l@{\hskip 10pt}l}
\toprule
 & & \multicolumn{6}{c}{$k=20$} && \multicolumn{6}{c}{$k=50$}  \\ \cmidrule{3-8} \cmidrule{10-15}
\multirow{2}{*}{Dataset} & \multirow{2}{*}{Method} & \multicolumn{2}{c@{\hskip 20pt}}{\textit{User Acc.}}     & \multicolumn{3}{c@{\hskip 20pt}}{\textit{Item Acc. (MDG)}} & \textit{Exposure} && \multicolumn{2}{c@{\hskip 20pt}}{\textit{User Acc.}}     & \multicolumn{3}{c@{\hskip 20pt}}{\textit{Item Acc. (MDG)}} & \textit{Exposure} \\
&  & NDCG & Recall & Min80\%      & Max5\%     & All     & Gini-Div.     && NDCG & Recall & Min80\%      & Max5\%     & All     & Gini-Div.                         \\ \midrule  

\multirow{7}{*}{Clothing}      & KNN         & 0.0738 & 0.1429  & 0.0270         & 0.3093 & 0.0604          & 0.6090          && 0.0825 & 0.1826 & 0.0349          & 0.3204 & 0.0689          & 0.5734             \\[0.9pt] 
      &  Heater      & 0.0549 & 0.1216 & 0.0095          & 0.2873 & 0.0389          & 0.3612          && 0.0686 & 0.1847 & 0.0193          & 0.3001 & 0.0507          & 0.4666         \\[-1.3pt] 
     &  Heater+MS    & 0.0554 & 0.1210 & \textbf{0.0127}$^\ast$ & 0.2766$^\dagger$ & 0.0412$^\ast$          & \textbf{0.5225}$^\ast$ && 0.0689 & 0.1835 & \textbf{0.0228}$^\ast$ & 0.2921$^\dagger$ & 0.0533$^\ast$          & \textbf{0.6152}$^\ast$  \\[0.2pt] 
     & GAR           & 0.0539 & 0.1145 & 0.0075          & 0.2921 & 0.0366          & 0.3573          && 0.0677 & 0.1785 & 0.0167          & 0.3073 & 0.0482          & 0.4643        \\[-1.3pt]
     & GAR+MS        & 0.0541 & 0.1148 & \textbf{0.0117}$^\ast$ & 0.2785$^\dagger$ & 0.0402$^\ast$          & \textbf{0.5930}$^\ast$ && 0.0675 & 0.1771$^\dagger$ & \textbf{0.0213}$^\ast$ & 0.2925$^\dagger$ & 0.0517$^\ast$          & \textbf{0.6773}$^\ast$  \\[0.2pt]
     & GoRec         & 0.0640 & 0.1394 & 0.0177          & 0.3065 & 0.0499          & 0.4832          && 0.0777 & 0.2028 & 0.0283          & 0.3200 & 0.0619          & 0.5626        \\[-1.3pt] 
     & GoRec+MS      & 0.0639 & 0.1371 & \textbf{0.0205}$^\ast$ & 0.2973$^\dagger$ & 0.0520$^\ast$          & \textbf{0.6431}$^\ast$ && 0.0776 & 0.2001$^\dagger$ & \textbf{0.0312}$^\ast$ & 0.3114$^\dagger$ & 0.0639$^\ast$          & \textbf{0.6904}$^\ast$  \\[0.9pt] 
     \hline \\[-1.0em]
\multirow{7}{*}{Electronics} & KNN           & 0.0188 & 0.0362 & 0.0030          & 0.1536 & 0.0179          & 0.4313          && 0.0215 & 0.0474 & 0.0051          & 0.1622 & 0.0209          & 0.3350          \\[0.9pt] 
    &  Heater    & 0.0144 & 0.0324 & 0.0000          & 0.0986 & 0.0070          & 0.1355 && 0.0206 & 0.0605 & 0.0005          & 0.1194 & 0.0109          & 0.2044           \\[-1.3pt] 
    &  Heater+MS & 0.0132$^\dagger$ & 0.0298$^\dagger$ & \textbf{0.0003}$^\ast$ & 0.1011 & \textbf{0.0092}$^\ast$ & \textbf{0.4821}$^\ast$          && 0.0188$^\dagger$ & 0.0553$^\dagger$ & \textbf{0.0025}$^\ast$ & 0.1173 & \textbf{0.0137}$^\ast$ & \textbf{0.5635}$^\ast$         \\[0.2pt] 
    & GAR        & 0.0150 & 0.0326 & 0.0000          & 0.1015 & 0.0074          & 0.2460          && 0.0206 & 0.0578 & 0.0008          & 0.1194 & 0.0112          & 0.3443         \\[-1.3pt]
    & GAR+MS     & 0.0131$^\dagger$ & 0.0289$^\dagger$ & \textbf{0.0001}$^\ast$ & 0.1003 & \textbf{0.0084}$^\ast$ & \textbf{0.5030}$^\ast$ && 0.0181$^\dagger$ & 0.0519$^\dagger$ & \textbf{0.0016}$^\ast$ & 0.1153$^\dagger$ & 0.0122$^\ast$          & \textbf{0.5773}$^\ast$         \\[0.2pt]
    & GoRec      & 0.0175 & 0.0377 & 0.0000          & 0.1212 & 0.0092          & 0.1496          && 0.0241 & 0.0675 & 0.0013          & 0.1420 & 0.0138          & 0.2159         \\[-1.3pt] 
    & GoRec+MS   & 0.0166$^\dagger$ & 0.0359$^\dagger$ & \textbf{0.0004}$^\ast$ & 0.1210 & \textbf{0.0110}$^\ast$ & \textbf{0.3254}$^\ast$ && 0.0228$^\dagger$ & 0.0641$^\dagger$ & \textbf{0.0027}$^\ast$ & 0.1392$^\dagger$ & \textbf{0.0158}$^\ast$ & \textbf{0.3836}$^\ast$         \\[0.9pt] \hline \\[-1.0em]
\multirow{7}{*}{Microlens}           & KNN   & 0.0567 & 0.1143 & 0.0171          & 0.3287 & 0.0511          & 0.5340          && 0.0657 & 0.1554 & 0.0242          & 0.3405 & 0.0597          & 0.5013         \\[0.9pt] 
    &  Heater    & 0.0500 & 0.1186 & 0.0073          & 0.3266 & 0.0386          & 0.3255          && 0.0672 & 0.1983 & 0.0173          & 0.3426 & 0.0523          & 0.4293          \\[-1.3pt]
    &  Heater+MS & 0.0517 & 0.1202 & \textbf{0.0113}$^\ast$ & 0.3106$^\dagger$ & 0.0420$^\ast$          & \textbf{0.5711}$^\ast$ && 0.0687 & 0.1989 & \textbf{0.0221}$^\ast$ & 0.3286 & 0.0560$^\ast$          & \textbf{0.6491}$^\ast$         \\[0.2pt] 
    & GAR        & 0.0470 & 0.1124 & 0.0069          & 0.3034 & 0.0356          & 0.4160          && 0.0647 & 0.1938 & 0.0162          & 0.3211 & 0.0491          & 0.5299         \\[-1.3pt]
    & GAR+MS     & 0.0473 & 0.1114 & \textbf{0.0092}$^\ast$ & 0.2904$^\dagger$ & 0.0369$^\ast$          & \textbf{0.6223}$^\ast$ && 0.0644 & 0.1901 & \textbf{0.0187}$^\ast$ & 0.3096$^\dagger$ & 0.0503$^\ast$          & \textbf{0.7085}$^\ast$         \\[0.2pt]
    & GoRec      & 0.0564 & 0.1284 & 0.0084          & 0.3563 & 0.0445          & 0.3208          && 0.0768 & 0.2217 & 0.0190          & 0.3729 & 0.0594          & 0.4078         \\[-1.3pt]
    & GoRec+MS   & 0.0574$^\ast$ & 0.1302 & \textbf{0.0119}$^\ast$ & 0.3412$^\dagger$ & 0.0470$^\ast$          & \textbf{0.4936}$^\ast$ && 0.0771 & 0.2204 & \textbf{0.0226}$^\ast$ & 0.3592$^\dagger$ & 0.0617$^\ast$          & \textbf{0.5613}$^\ast$        \\ 
\bottomrule 
\end{tabular}
\end{table*}

\subsection{Mitigation}
Given the above analysis, we propose a simple post-processing method to mitigate this biased predictive behavior. In the typical popularity bias context, post-processing approaches often use training set item popularity to determine which items to promote in user rankings \cite{abdollahpouri2021user, zhu2021popularity}. However, this information is not available for cold items, and, as we saw in Section \ref{sec:motiv}, it is often not the most popular cold items that are overexposed. Instead we note that the learned embeddings of popular items tend to have large vector magnitudes, as observed in previous works \cite{ren2022mitigating, allen2022mitigating, chen2023adap, kim2023test, lee2024post}. Theoretical analyses \cite{ren2022mitigating, chen2023adap} show that the vector magnitudes of embeddings for popular items will grow more rapidly than those of items in the long-tail due to their more frequent sampling during training. Since user-item preference scores are calculated by dot products, these higher magnitudes force items higher in user rankings even when similarity in vector direction is fairly low, leading to biased predictive behavior as the model converges. 

As shown in Figure \ref{fig:mag}, a similar correlation between magnitude and prediction count also holds for inferred popularity in the content-based predictions of our cold models. We therefore propose to balance model predictions by scaling the magnitudes of output item representations. While a similar idea has previously been applied in warm contexts \cite{kim2023test,lee2024post}, these works either use a global normalization factor \cite{kim2023test} or leverage the available popularity information to dynamically adjust the degree of scaling based on item popularity \cite{lee2024post}. To make such a technique suitable for the cold item scenario, we adapt our approach based on the above insight that embedding magnitude is a proxy for predicted popularity. To ensure that the prevalence of overexposed items is reduced, we want the impact of the scaling to be higher for more extreme vector magnitudes. We also want to adjust the influence of the magnitude in a controllable way, so that model accuracy can be balanced with item fairness. To accomplish these goals, our proposed method modifies vector magnitudes such that the standard deviation of the magnitude distribution is linearly scaled while the mean is held constant. In other words, given a cold item $c$ and its output vector $x_c$, we define our scaling factor $\gamma_c$ such that
\begin{equation}
    ||\gamma_c x_c||-\mu_w = \frac{||x_c||-\mu_w}{1+\alpha}
\end{equation}
where $||\cdot||$ denotes vector magnitude, $\mu_w$ is the mean magnitude of the warm item vector outputs, and hyperparameter $\alpha>0$ controls the strength of the scaling. We solve for $\gamma_c$ to get
\begin{equation}
    \gamma_c = \left(\frac{||x_c||+\alpha \mu_w}{||x_c||(1+\alpha)}\right)
\end{equation}
We can see that this method achieves our two aims: the impact of the scaling increases when $||x_c||$ is further from $u_w$, and $||\gamma_c x_c||$ will tend towards $\mu_w$ as $\alpha$ grows, i.e. the vectors will converge to being normalized (with an additional constant factor of $\mu_w$, which will not affect model rankings). This approach thus provides controllability between the original unscaled magnitudes and full vector normalization, as desired. Notably, unlike previous methods \cite{kim2023test,lee2024post}, this convergence towards $\mu_w$ happens from both tails of the magnitude distribution: smaller vector magnitudes are pulled up towards the mean while the large outliers are pulled down, rather than all magnitudes shrinking towards 1. This improves the balancing effect for intermediate values of $\alpha$, as the increased prevalence of rarer items and reduced exposure of overrepresented items are achieved simultaneously. As we show in Section~\ref{sec:results}, calculating preference scores via dot product with these scaled vectors significantly reduces the influence of predicted popularity in item rankings.

\begin{figure*}[]
\begin{tabular}{ccc}
    \includegraphics[width=54mm]{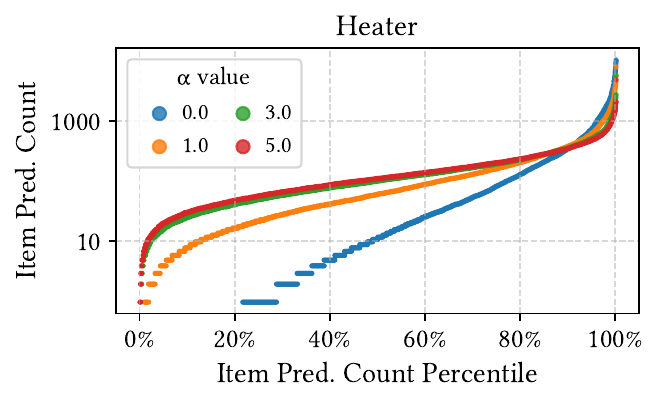} & 
     \includegraphics[width=54mm]{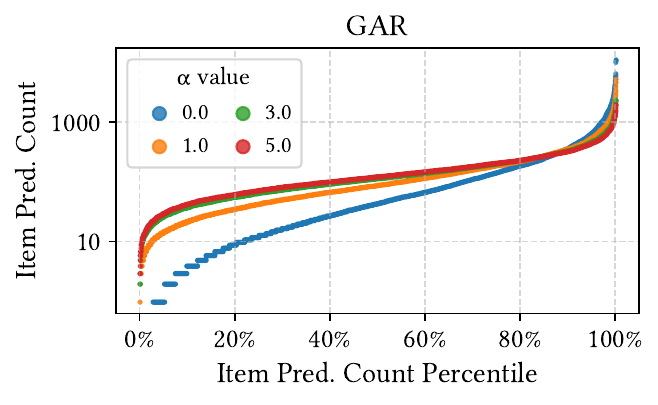} & \includegraphics[width=54mm]{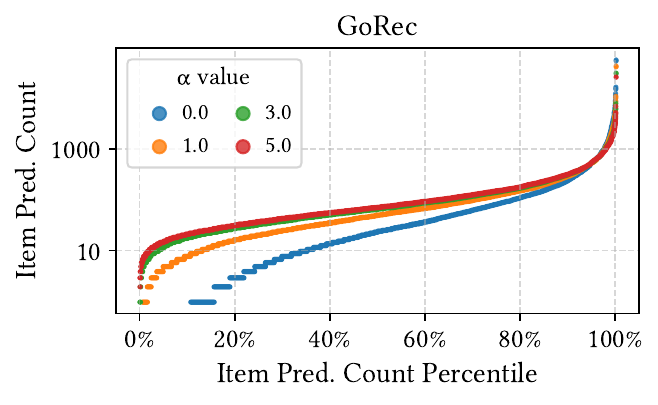} \\ 
\end{tabular}
\caption{Cold test set item prediction counts at $k=20$ against item prediction count percentiles (i.e.\ each item's position in the sorted list of prediction counts) in the Electronics dataset. Only items predicted at least once are plotted.} \label{alpha_pred}
\Description{Prediction counts against prediction count percentiles for three models with magnitude scaling. This figure contains three scatter plots corresponding to the Heater, GAR, and GoRec models. Each plot contains four sets of points corresponding to alpha values of 0.0, 1.0, 3.0, and 5.0, all in different colors. The y-axis is item prediction counts on a log-scale, and the x-axis is item prediction count percentile from 0 percent to 100 percent. In all plots, the 0.0 alpha points show a much steeper descent from the max value in the top right to the intersection with the x-axis. The points with nonzero alpha intersect the axis at or very near 0 percent, while the 0.0 alpha points intersect at near 20 percent for Heater, 5 percent for GAR, and 10 percent for GoRec. There is also a visible gap above the 0.0 alpha points and below the rest between 0 percent and approximately 80 percent in all plots. There is less difference between the points with nonzero alpha, but the curves increasingly flatten and converge as alpha increases.}
\end{figure*}

\section{Results and Discussion}\label{sec:results}
We report model performance pre- and post-magnitude scaling~(\textbf{MS}) across all datasets in Table \ref{tab:full_results}, averaging results on the cold test set over five runs at the optimal validation set hyperparameters, with ranking cutoffs $k$ of 20 and 50. We tune the scaling parameter $\alpha$ in the range $\{0.5,1.0,1.5,2.0,2.5,3.0,4.0,5.0\}$ on the validation set, selecting values that balance the trade-off discussed in \cite{zhu2021fairness} between item fairness and loss of performance in user-level accuracy. 

We observe first that our magnitude scaling method significantly boosts prediction diversity (Gini-Div.) across all configurations. Concretely, this means that top ranking positions are distributed more equitably among items, as illustrated in Figure \ref{alpha_pred}. Our magnitude scaling increasingly balances the distribution of prediction counts as $\alpha$ grows, with the curves converging towards the limit where all vectors have a magnitude of $\mu_w$. While the unscaled models leave up to 20\% of items out of the top 20 predictions for all users, the scaled versions provide at least 10 predictions to all but a few items. Returning to the GoRec Electronics case from Section \ref{sec:motiv}, the scaling reduces the proportion of predictions for the top 50 items from 27.6\% to 13.6\% and the number of entirely unrepresented items from 1,326 to just 6. Further analysis of the impact of $\alpha$ and the trade-off between accuracy and fairness across all datasets can be found in the supplementary material.

We can see in Table \ref{tab:full_results} that this increased exposure also helps to provide statistically significant improvements to low-end item accuracy, with MDG-Min80\% typically increasing by at least 20\% when $k=20$ and at least 10\% when $k=50$. The balancing process does slightly reduce top-end MDG, but still results in net improvement to MDG-All. Its cost to user accuracy is also small: results stay consistent or improve slightly for Clothing and Microlens, and only decrease slightly for Electronics (typically by less than 10\%). 

These findings illustrate the unique impact of popularity bias in cold item contexts. In biased warm models, overexposed items are very popular, and since predictions on these items are more likely to be correct, reducing their prevalence will likely also reduce the rate of user-item hits. However, as seen in Section \ref{sec:motiv}, predicted cold item popularity often does not match actual audience preferences. This means that, rather than removing correct but `easy' predictions, dampening model predictive biases can free up high ranking spots for more relevant items without much damage to overall accuracy.

Finally, we note that the baseline KNN method often outperforms the cold-start models in both accuracy and fairness, particularly in the metrics sensitive to predicted item rank and when $k=20$. This suggests that KNN's strength is in placing relevant items in the highest ranking positions, while its more balanced item performance illustrates that it is less vulnerable to predictive biases based on interaction history. As shown by our other findings, there are significant difficulties in using item content alone to directly predict collaborative behavior, so in future work we plan to explore whether these benefits of content similarity-based inference can be leveraged further to improve outcomes for both users and items.

\section{Conclusion}
In this paper, we examine the effects of warm model popularity bias on generative cold-start training. We show via a detailed case study that these cold-start models inherit the biases of their supervisory warm model, overexposing certain items due to similarity in content with popular items from the training set. Noting that the representations for these overexposed items often have large vector magnitudes, we demonstrate via comprehensive evaluation on three real-world datasets that the effects of this bias can be mitigated by compressing the vector magnitude distribution, thereby increasing diversity and fairness while largely preserving or improving accuracy. As with all post-processing bias mitigation methods, this approach treats the symptoms of inherited predictive bias rather than the cause, and future work will investigate the application of in-processing mitigation techniques to this problem, e.g. by regularizing learned vector magnitudes \cite{allen2022mitigating} during cold-start training. 

\begin{acks}
This work was funded by UKRI as part of the UKRI CDT in Artificial Intelligence and Music [grant number EP/S022694/1].
\end{acks}
\bibliographystyle{ACM-Reference-Format}
\bibliography{bibfile}

\end{document}